\documentclass[prl,twocolumn,superscriptaddress]{revtex4-1}
\usepackage{amsmath,amssymb,bm}
\newcommand{\braket}[1]{\langle {#1} \rangle }
\newcommand{\ket}[1]{|{#1} \rangle }

\usepackage{amsmath}	
\usepackage{color}
\usepackage{txfonts}
\usepackage[pdftex]{graphicx}
\usepackage{hyperref}

\begin{document}
		\title{Quantum entanglement in nuclear Cooper pair tunneling with $\gamma$-rays}
	\author{G. Potel}
	\affiliation{Lawrence Livermore National Laboratory, Livermore, California 94550, USA}	
	\author{F. Barranco}
	\affiliation{Departamento de F\'isica Aplicada III, Escuela Superior de Ingenieros, Universidad de Sevilla, Camino de los Descubrimientos, Sevilla, Spain}
	\author{E. Vigezzi}
	\affiliation{INFN Sezione di Milano, Via Celoria 16, I-20133 Milano, Italy}
	\author{R. A. Broglia}
	\affiliation{The Niels Bohr Institute, University of Copenhagen, DK-2100 Copenhagen, Blegdamsvej 17, Denmark
		}
\affiliation{Dipertimento di Fisica, Universit\`a degli Studi di Milano, Via Celoria 16, I-20133 Milano, Italy}
	\date{\today}
	\begin{abstract}
While Josephson-like junctions, transiently established in heavy ion collisions ($\tau_{coll}\approx10^{-21}$ s) between superfluid nuclei --through which Cooper pair tunneling ($Q$-value $Q_{2n}$) proceeds mainly in terms of successive transfer of entangled nucleons-- is deprived from the macroscopic aspects of a supercurrent, it displays many of the special effects associated with spontaneous symmetry breaking in gauge space (BCS condensation), which can be studied in terms of individual quantum states and of tunneling of single Cooper pairs. From the results of studies of one- and two- neutron transfer reactions carried out at energies below the Coulomb barrier we estimate  the value of the mean square radius (correlation length) of the nuclear Cooper pair. A quantity related to the largest distance of closest approach  for which the  absolute two-nucleon  tunneling cross section is of the order of the single-particle one.  Furthermore,  emission of $\gamma$-rays of (Josephson) frequency $\nu_J=Q_{2n}/h$ distributed over an energy range  $\hbar/\tau_{coll}$ is predicted.
	\end{abstract}	
	\maketitle	
\textit{Introduction}---	A pair of interacting electrons moving in time-reversal states $(\nu,\tilde \nu)(\equiv(\mathbf k\uparrow,-\mathbf k \downarrow))$ above a  non-interacting Fermi sea whose only role is to block, through Pauli principle, states below the Fermi energy $\epsilon_F$ from participating in the two-particle system, lead to   a bound state provided the interaction is attractive, no matter how weak it is \cite{Cooper:56}.
	
At the basis of BCS superconductivity \cite{Bardeen:57a,Bardeen:57b} one finds the condensation of strongly overlapping, very extended, weakly bound Cooper pairs corresponding to ordering in occupying momentum space, and not space-like condensation of strongly bound clusters which undergo Bose condensation. In BCS condensation, the inner, intrinsic structure of the pair, that is, the fact that it is made out of fermions entangled in time reversal states, is the characterizing feature, with its energy gap for both single-pair translation and dissociation (see \cite{Penrose:51,Penrose:56,Yang:62,Anderson:96}), as it emerges from Schrieffer's trial wavefunction $\ket{\Psi_{BCS}}=\prod_{\nu>0}(U'_\nu+e^{-2i\phi}V'_\nu P^\dagger_\nu)\ket{0}$ \cite{Schrieffer:64}.	
	The associated spontaneously broken symmetry in the two-dimensional gauge space, is quantitatively measured by the generalized deformation (order) parameter
	 $\alpha_0=\braket{\Psi_{BCS}|P^\dagger|\Psi_{BCS}}=e^{-2i\phi}\alpha_0'$.
	  The pair creation operator is defined as $P^\dagger=\sum_{\nu>0}a^\dagger_\nu a^\dagger_{\tilde\nu}$, where $a^\dagger_\nu (a^\dagger_{\tilde\nu})$ creates, acting on the vacuum state $\ket{0}$, a fermion (electron) moving in the state $\nu (\tilde\nu)$ while $\alpha'_0=\sum_{\nu>0}U'_\nu V'_\nu$ measures the number of Cooper pairs, a quantity closely related to the pairing gap $\Delta'=G\alpha_0'$ ($\approx1$ meV), $G$ being the pairing coupling constant. The intrinsic, body-fixed frame of reference ($x'$-axis) subtends a gauge angle $2\phi$ with the laboratory axis $x$ (see e.g. Fig. 11 ref. \cite{Potel:13b}).
	  
	\emph{Weakly coupled superconductors}---
The Cooper pair  wavefunction can be written as $\braket{\mathbf r_1\sigma_1,\mathbf r_2\sigma_2|\sum_{\nu>0}c'_\nu P^\dagger_\nu|0}=\varphi_q(\mathbf r)e^{i\mathbf q\cdot\mathbf R}\chi(\sigma_1,\sigma_2)$. The variable $\mathbf r\,(\mathbf R)$ is the relative (center of mass) coordinate while $\mathbf q$ is the center of mass momentum, $\chi$ being the singlet spin function. For $q=0$ \cite{Schrieffer:64,Weisskopf:81}
\begin{align}\label{eq100}
\varphi_0(\mathbf r)\sim e^{-r/\xi}\cos k_F r,
\end{align}
the quantity
\begin{align}\label{eq101}
\xi=\frac{\hbar  v_F}{\pi\Delta},
\end{align}
being the correlation length ($\approx10^4$\AA).

 In the calculation of the Cooper pair tunneling probability $p_2$ between two \textit{weakly coupled superconductors (\textbf{S}-\textbf{S}), also known as Josephson junction}, of typical width $d(\approx10-30\text{ \AA}\ll\xi)$ one has to add the phased amplitudes before one takes the modulus squared. As a consequence, the probability $p_2$ of a pair going through the junction is comparable to the probability $p_1$ for a single electron (\cite{Pippard:12}, see also Ch. 6 \cite{Tinkham:96}). 
	This result is at the basis of the Josephson effect(s) \cite{Josephson:62,Josephson:73,Anderson:63,Anderson:64b,Shapiro:63}: \textbf{a}) \textit{unbiased junction}; the small but finite overlap of the condensed amplitudes $\ket{\Psi_{BCS}(\ell)}$ and $\ket{\Psi_{BCS}(r)}$ is sufficient to lock the associated gauge phase difference $(\phi_{rel}(\mathbf R)=\phi_\ell-\phi_r)$, function which acts as the velocity potential of a collective flow (center of mass momentum) superimposed on the Cooper pairs correlated intrinsic motion (\ref{eq100}). The associated direct supercurrent of carriers of charge $q=2e$ and maximum  value $I_c=\frac{\pi}{e}\Delta_{\ell r}\frac{1}{R_b}$ is undamped, because the internal degrees of freedom are frozen by the reduced pairing gap $\Delta_{\ell r}=\frac{\Delta_\ell\Delta_r}{\Delta_\ell+\Delta_r}$. In the above relation $\Delta_\ell$ and $\Delta_r$ are the pairing gap of the left and right superconductors with respect to the junction. Similarly concerning the gauge phase $\phi_\ell$ and $\phi_r$; \textbf{b}) \textit{biased junction}; when there is a dc voltage $V$, and thus an associated chemical potential difference ($\lambda_{\ell}-\lambda_r$) across the junction, circulation of an alternating current  of carriers $q=2e$, critical value $I_c$ and of frequency $\nu_J=V\times2e/h$ is observed, while  $\phi_{rel}$ precesses at the rate given by $\dot\phi_{rel}=(\lambda_\ell-\lambda_r)/\hbar=V\times2e/\hbar$.  There is then an energy difference $\Delta E=V\times2e$ each time a Cooper pair tunnels from one   side of the junction to the other, energy which must appear elsewhere. Being the process superconducting, it is  free of dissipation. To leave the quasiparticle distribution unchanged, Cooper pairs can  tunnel back and forth with the emission of a photon of frequency $\nu_J$. The Josephson junction not only converts a direct voltage into an alternating current, but also works as an oscillatory circuit. It radiates electromagnetic waves in the superhigh frequency range.

	The critical supercurrent  $I_c$ (of typical value $\approx 2$ mA) across an \textbf{S}-\textbf{S} junction is, within a factor $\pi/4$ equal to the \textbf{N}-\textbf{N} single electron carrier current, for an applied equivalent potential bias $V_{eq}=(2\Delta/e)\, (\approx 2$ mV), S (N) indicating the superconducting  (normal) phase of the metal. That is
		\begin{align}\label{eq2}
		I_c=\frac{\pi}{4}\frac{V_{eq}}{R_b},
	\end{align}	
	where $R_b(\approx1\Omega)$ is the  resistance of the junction ($I_c\approx1.6$ mA). A relation which testifies to the correctness of $p_2\approx p_1$, and constitutes one of the pillars on which the validity of the BCS description of superconductivity rests. Another  one is provided by the photons of frequency 
	\begin{align}\label{eq3}
	\nu_J=K_J V,
	\end{align}
	emitted by a biased S-S junction. The Josephson constant, inverse of the flux quantum (fluxon) is  $K_J=2e/h$. For voltage differences across the junction of $\approx$1mV one has $\nu_J\approx0.5$ THz. 
	
	It is of notice that in the tunneling process between two superconductors in which a bias of value $V\gtrsim2\Delta/e$ is applied to the junction,  Cooper pairs are broken and quasiparticle excitations created --thus the labeling (\textbf{S}-\textbf{Q}) given in the literature to such processes-- through which a normal (dissipative) current of carriers $q=e$ flows \cite{Giaver:73}. In other words, for $T=0$ one is in presence of processes connecting a ground state (S) with ground and excited states (Q). The importance of this fact in connection with the Josephson-like junction transiently formed in heavy ion reactions between superfluid nuclei, becomes apparent below. 	
	
	\emph{Cooper pair tunneling in nuclei}--- 
	Recently, one- and two- neutron transfer reactions between superfluid nuclei have been studied making use of magnetic and gamma spectrometers with heavy ion reactions in inverse \cite{Montanari:14}  and direct \cite{Montanari:16} kinematics,
		\begin{align}\label{eq4}
		^{116}\text{Sn}+^{60}\text{Ni}\to\left\{\begin{array}{c}
		^{115}\text{Sn}+^{61}\text{Ni}\quad (Q_{1n}\approx -1.74\text{ MeV}),\quad\quad\text{(a)} \\ [10pt]
		^{114}\text{Sn}+^{62}\text{Ni}\quad (Q_{2n}\approx 1.307\text{ MeV}).\quad\quad\text{(b)}
		\end{array} \right.
		\end{align}
These reactions were carried out for twelve bombarding energies in the range 140.60 MeV$\leq E_{cm}\leq167.95$ MeV. That is, from energies above the Coulomb barrier ($E_B=157.60$ MeV), to well below it. While the Cooper pair transfer channel (\ref{eq4} (b)) is dominated by the ground-ground state transition, the single-particle transfer one is inclusive. In fact, the theoretical calculations of the differential cross section associated with channel (\ref{eq4} (a)) indicates the incoherent contribution of 10 quasiparticle states of $^{61}$Ni lying at energies $\lesssim 2.640$ MeV (see Table 1 of reference \cite{Montanari:14} as well as reference \cite{Lee:09}). A value  which is consistent with twice  the value of the pairing gap of Ni. In other words,  in the case of the reaction \ref{eq4} (a), we are in presence of a  S-Q like transfer\footnote{Within this context one observes in e.g. the reaction $^{116}$Sn$(p,t)^{114}$Sn, $\sigma(gs\to gs)=2492 \mu$b, a cross section to be compared with $\Sigma_i\sigma(gs\to2qp(i))=2012 \mu$b (1.3 MeV$\leq E_{2qp}(i)\leq4.14$ MeV, Table I \cite{Guazzoni:04}), i.e. $\sigma(gs-2qp)/\sigma(gs-gs)\approx0.81$, a quantity which comes  close to $(\pi/4)^2$ (see last paragraph  before Eq. (\ref{eq2})).}. Making use of the relation (\ref{eq101}), as well as of the values ($v_F/c)\approx 0.3$ and $\Delta\approx1.3-1.5$ MeV, one obtains $\xi\approx13.6$ fm (within this context see for example Fig. 7, App. A of \cite{Potel:17}).

The analysis of the data associated with the reactions (\ref{eq4} (a)) and (\ref{eq4} (b))  carried out in \cite{Montanari:14,Montanari:16} makes use of a powerful semiclassical approximation in which  the optical potential employed was microscopically calculated. The short wavelength of relative motion (de Broglie reduced wavelength $\lambdabar=0.36/2\pi$ fm$\approx0.06$ fm), allows to accurately determine the distance of closest approach for each bombarding energy. Making use of the $U,V$ occupation amplitudes for both Sn and Ni, as well as the optical potential given in \cite{Montanari:14} (see also \cite{Broglia:04a}), we have calculated, within the framework of first and second order DWBA \cite{Potel:13}, the absolute one- and two- nucleon transfer differential cross sections.  In the second case, including both successive (dominant channel) and simultaneous transfer, properly corrected by non-orthogonality. Theory is compared with experiment in Table  \ref{tab1}. As expected \cite{Montanari:14}, the results  provide an overall account of the experimental findings.

 From direct inspection of this Table it emerges that the distance of closest approach lying within the interval 13.12 fm $\leq D_0 \leq13.49$ fm is the largest one for which $d\sigma /d\Omega|_{2n}$ is, within a factor of about 0.6 ($\approx\left(\pi/4\right)^2$) of the same order of $d\sigma /d\Omega|_{1n}$. In keeping with (\ref{eq100}) and (\ref{eq101}) one can posit that the above interval  provides a sensible bound to the size of the nuclear Cooper pair correlation length. Already increasing $D_0$ by $\approx$0.6 fm ($D_0=14.05$) $\sigma_{1n}$ becomes a factor 6 larger than $\sigma_{2n}$. A  signal indicating that stretching the transferred Cooper pair to larger dimensions ruptures it, quenches its pairing gap and unfreezes the quasiparticle degrees of freedom. Said it differently,  a consequence  of forcing Cooper pair partners, in the dominant successive transfer process, to be at a relative distance  larger than $\xi$.  This leads to a strain which plays a role similar to that played by applying a momentum $q\approx1/\xi$ (associated with the critical bias $V_{eq}=2\Delta/e$) to the center of mass of the Cooper pairs,  resulting in the  transition from the S-S transfer regime to the S-Q one. As a result, we choose $D_0=13.49$ fm as a representative value for $\xi$ of the transferred Cooper pair.
 
\emph{Nuclear analogue of radiating Josephson junction}---
As stated before, when the two superconducting elements of a junction are at a different electric potential, the transfer of a pair of electrons from one side (e.g. $\ell$) to the other one ($r$) involves an energy change of $2e\times V$. If the process is truly a superfluid process, free of dissipation, this energy must appear elsewhere as a unit. In fact, it appears as a photon of energy $h\nu=2e\times V$ (radiofrequency) in keeping with (\ref{eq3}), and as experimentally observed (see e.g. \cite{Lindelof:81} and refs. therein).

In the nuclear case and in connection with the reaction \ref{eq4} (b) for bombarding conditions for which $D_0=13.49$ fm (namely $E_{c.m.}\approx154.26$ MeV and $\tau_{coll}\approx\xi/(2E_{c.m.}/\mu_i)^{1/2}\approx 0.5\times10^{-21}$s), the Cooper pair tunnels few MeV below the Coulomb barrier. Consequently the absorptive component of the optical potential plays essentially no role in the process, and tunneling takes place lossless, free of dissipation. Being the bombarding energy $\approx3.9$ MeV/$A$ ($E_{lab}=452.5$ MeV), that is an order of magnitude smaller than the Fermi energy, one can expect that there can be time for the two-neutrons to be transferred back and forth about three times. That is, for about two ($\approx1.5$) cycles of the quasielastic process 
\begin{align}\label{eq8}
^{116}\text{Sn}+^{60}\text{Ni}\to^{114}\text{Sn}+^{62}\text{Ni}\to^{116}\text{Sn}+^{60}\text{Ni}.
\end{align}
	Due to the fact that nuclear Cooper pairs carry an effective charge $(e)_{eff}\approx(-2eZ/A)$, one expects the transient Josephson-like nuclear junction to emit $\gamma$-rays of frequency $\nu=Q_{2n}/h$ (=1.307 MeV/$h$). Because of the short collision time ($\tau_{coll}$) the radiated photons will display  a width ($\approx\hbar/\tau_{coll}$). Due to the  recoil of the $\ell$(Sn)-$r$(Ni) nuclear superconducting junction associated with  Cooper pair tunneling, the corresponding line shape will be distorted with respect to a Gaussian-like shape. A recognition of the fact that after all, the nucleus is not an infinite, but a finite quantal many-body system. It, nonetheless, displays some important aspects of the macroscopic behavior of a Josephson junction, due to two main reasons. First, because of quantum coherence existing between the Cooper pairs of a superfluid nucleus. Second, due to entanglement over distances of the order of $\xi$ of the partner nucleons of each Cooper pair. In what follows we elaborate on these points, and calculate the associated $\gamma$-emission cross section in terms of a macroscopic formulation of the (ac) Josephson effect, particularly suited to be used in connection with the nuclear case.
	
	Concerning the search for nuclear analogs of the Josephson effect see (\cite{Goldanskii:68,Dietrich:70,Dietrich:71b,Hara:71,Kleber:71,Weiss:79,vonOertzen:01}, see also \cite{Broglia:04a}).

	\emph{Macroscopic calculation of dipole emission}---
	Making use of $\alpha_0=e^{-2i\phi}\alpha_0'$ one can introduce the density of superconducting electron (fermion) pairs,
		\begin{align}\label{eq9}
\Psi^*\Psi=\frac{\alpha_0'}{\mathcal V}=n_s',
\end{align}	
	in terms of the pair probability amplitude \cite{Ginzburg:50,Ginzburg:04},
\begin{align}\label{eq10}
\Psi=e^{-i\phi}\sqrt{n'_s},
	\end{align}	
where $\mathcal V$ is an appropriate volume element. Both $n'_s$ and $\phi$ can be functions of space (and time), and their variation determine the motion of the BCS condensate, e.g. the supercurrent. Since the pairs are in the same state and must therefore behave in an identical fashion, the equations of motion of the macrostate must coincide with the equation of motion for any single pair of this state \cite{Marcerau:69}. In other words, due to its unique coherence properties the condensed (superfluid) portion of the superconductor behaves like a single quantum particle of mass and charge twice that of an electron. 
	
	It is then sensible to expect that the dynamical behavior of a Josephson junction --right ($r$) and left ($\ell$) weakly coupled superconductors-- would be similar to that of two quantum levels weakly coupled to one another via an external field \cite{Feynman:63}. Considering the situation in which the tunneling interaction is relatively constant 
	over a coherence length \cite{Rogovin:76}, the electrodynamics of a radiating Josephson junction is analogous to that of a two-level atom placed in a static external field, role which in the present case is played by the tunneling interaction inducing non resonant transitions between the two quantum levels. These transitions give rise to an induced dipole moment whose oscillations generate the coherent Josephson radiation field, the intensity of the emitted radiation being proportional to the number of Cooper pairs that are involved in the tunneling process quantity squared, the frequency being that defined in Eq. (\ref{eq3}).
	
	Superradiance is a collective effect arising from an ensemble of two-level emitters locking, due to the presence of an external field, their dipole oscillations in phase, giving rise to an enhanced decay rate many times faster than the spontaneous emission rate of individual emitters, in quadratic proportion with the number of emitters.

	A similar, incipient superradiant Josephson-like phenomenon is expected to arise in the  case of the nuclear heavy ion reaction under discussion, from an ensemble of correlated Cooper pairs ($\alpha_0'\approx8$ (2),  $^{116}$Sn ($^{60}$Ni)) undergoing the coherent, back and forth quasielastic Cooper pair transfer process. In what follows the associated $\gamma$-emission probability is calculated. 
	
	According to Fermi's Golden rule, the rate of spontaneous emission between two levels in the dipole approximation can be written as (\cite{Basdevant:05}, p. 340),
	\begin{align}\label{eq12}
\frac{dP_{if}}{dt}=\frac{4\omega_{if}^3|\braket{i|\,\mathbf d\,|f}|^2}{3\hbar c^3},
	\end{align}	
	where $\omega_{if}=2\pi/\mathcal T$ is the emission frequency,  $\mathcal T$ being the associated period, $\mathbf d=q\mathbf r$ the dipole moment operator and $q$ the  charge.
	
	In connection with the reaction (\ref{eq4} (b))  $i\equiv B(=A+2)+b\to f\equiv A+a(=b+2)$, and $q=2e_{eff}=-2\times e(Z_b+Z_B)/(A_b+A_B)$,  where $(A_b,Z_b)\equiv(60,28)$ and $(A_B,Z_B)\equiv(116,50)$, one obtains  $q=-2\times(78/176)e=-e\times0.89$, and $d=-e\times0.89\times13.49$ fm=$-e\times12.01$ fm. 
	In keeping with the discussion on superradiance carried out above, $\frac{dP_{if}}{dt}=\mathcal N/\mathcal T$, $\mathcal N$ being the number of photons emitted per cycle,
\begin{align}\label{eq13}
\mathcal N=	\mathcal T\times \frac{dP_{if}}{dt}=\frac{8\pi}{3}\frac{(\hbar\omega_{if})^2d^2}{(\hbar c)^3}\approx3.71\times10^{-4},
\end{align}	
	where $\hbar\omega_{if}=Q_{2n}=1.307$ MeV. Making use of the experimental value (see Tab. \ref{tab1}), $d\sigma_{2n}(E_{cm}=154.26\text{MeV})/d\Omega|_{\theta_{c.m.}=140^\circ}\approx2.58$ mb/sr, one expects the associated $\gamma$-radiation to be emitted perpendicular to the reaction plane, with a cross section  $d\sigma_\gamma/(d\Omega_\gamma dE_\gamma)=\mathcal N d\sigma_{2n}/d\Omega|_{\theta_{c.m.}=140^\circ}\,\delta(E_\gamma-Q_{2n})\approx0.96\, \mu$b/sr $\delta(E_\gamma-Q_{2n})$. In other words, one expects a (reduced) strength function of centroid 1.307 MeV, width $\hbar/\tau_{coll}\approx1.3$ MeV and energy integrated area of 0.96 $\mu$b/sr.

	\emph{Dipole radiation: microscopic calculation}---
	A similar calculation of the $\gamma$-emission quasielastic process, this time fully microscopic, was carried out extending the second order DWBA formalism employed in the calculation of the two-nucleon transfer absolute differential cross sections displayed in Tab. \ref{tab1}, to include the coupling to the electromagnetic field in the dipole approximation.

	The $T$-matrix associated with the successive transfer of the Cooper pairs, that is, half of a cycle of the process leading to the result  (\ref{eq13}), and which contributes essentially all of the corresponding  cross section, can be written in the post-post representation as
	\begin{widetext}
		\begin{align}\label{eqgamma}
		\nonumber T&=2\sum_{\nu,\nu'}B^{(A)}_{\nu}B^{(b)}_{\nu'}\int \chi_f^*(\mathbf r_{Bb},\mathbf k_f)\left[\phi_{j_f}(\mathbf r_{A_1})\phi_{j_f}(\mathbf r_{A_2})\right]^{0*}_0 U^{(A)}(\mathbf r_{b1})\left[\phi_{j_f}(\mathbf r_{A_2})\phi_{j_i}(\mathbf r_{b_1})\right]^{K}_M d^1_{m_\gamma}(\mathbf r_{O1})\, d\mathbf r_{Cc}\, d\mathbf r_{b_1}\, d\mathbf r_{A_2}\\
		&\times\int G(\mathbf r_{Cc},\mathbf r'_{Cc})\left[\phi_{j_f}(\mathbf r'_{A_2})\phi_{j_i}(\mathbf r'_{b_1})\right]^{K*}_M U^{(A)}(r'_{c2})\left[\phi_{j_i}(\mathbf r'_{b_2})\phi_{j_i}(\mathbf r'_{b_1})\right]^{0}_0\chi_i(\mathbf r'_{Aa},\mathbf k_i)\, d\mathbf r'_{cC}\, d\mathbf r'_{b_1}\, d\mathbf r'_{A_2}.
		\end{align}
		\end{widetext}
			
where $B^{(i)}_j=\left(\sqrt{j+\tfrac{1}{2}}\,U^{(i)}_jV^{(i)}_j\right)$ is the two-nucleon transfer spectroscopic amplitude (see e.g. \cite{Potel:13}; see also \cite{Josephson:62}), while $U^{(A)}(r)$ is the mean field potential \footnote{It is of notice that the mean field potentials $U^{(A)}$ and $U^{(b)}$ are those used in the calculation of the single-particle wavefunctions appearing in Eq. (\ref{eqgamma}).} mediating the successive transfer process $B(=A+2)+b\to F(=A+1)+f(=b+1)\to A+a(=b+2)$. The Green's function $G(\mathbf r_{Cc},\mathbf r'_{cC})$
propagates the intermediate channel $(F,f)$ (no asymptotic waves), while $\chi_i,\chi_f$ are the distorted waves describing the relative motion of the heavy ions in the initial ($B,b$) and final $(A,a)$ channels, the momenta $k_i$ and $k_f$ ensuring energy conservation. The dipole operator is defined as
\begin{align}\label{eq7}
d^1_{m_\gamma}=q\sqrt{\frac{4\pi}{3}}\mathcal Y_{1m_\gamma}(\mathbf r_{O1})
\end{align}
where $\mathcal Y_{1m_\gamma}(\mathbf r_{O1})$ is the vector spherical harmonic of order one, $\mathcal Y_{1m_\gamma}(\mathbf r_{O1})=r_{O1}Y_{m_\gamma}^1(\hat r_{O1})$ and $r_{O1}$ is the coordinate of one of the transferred neutrons measured from the center of mass. Aside from this operator, the similitude of (\ref{eqgamma}) with the expression of the second order energy perturbation associated with the tunneling Hamiltonian of Cohen, Falicov and Phillips \cite{Cohen:62} (tunneling potential) acting between two superconductors (see also \cite{Anderson:64b})) is apparent. The first factor of (\ref{eqgamma}) involving the product of the two-nucleon transfer spectroscopic amplitudes is proportional to the product $\Delta_\ell\Delta_r$ (critical current), while the integral parallels the expression of the conductivity, i.e. the inverse of the junction resistance.

The $\gamma$-strength function (double differential cross section) associated with (\ref{eqgamma}) can be written as
  \begin{align}\label{eq156}
\nonumber\frac{d^2\sigma}{d\Omega dE_\gamma}&=\left(\frac{\mu_i\mu_f}{(2\pi\hbar^2)^2}\frac{k_f}{k_i}\right)\left(\frac{8\pi}{3}\frac{E_\gamma^2}{(\hbar c)^3}\right)\left|T_{m_\gamma}(\mathbf k_f,\mathbf k_i)\right|^2\\
&\times\delta(E_\gamma+E_f-(E_i+Q)),
\end{align} 
where $E_\gamma=\hbar\omega_{if}$, $k_i=(2\mu_iE_i)^{1/2}/\hbar$ and $k_f=(2\mu_fE_f)^{1/2}/\hbar$, $E_i$ and $E_f$ being the (c.m.) kinetic energy in initial and final channels.

In addition to the analytic prefactors describing the electromagnetic and kinematical phase spaces, the strength function (\ref{eq156}) depends on the photon energy through the distorted waves and the effective formfactors which, in channel ($F(A+1)$), $(f(=b+1))$, restrict the integrations to the region of overlap between the partner nucleons of the tunneling Cooper pair. In other words, for the overlap region associated with the largest relative distance between the two ions in which the normal and abnormal densities are simultaneously present. That is, the distance of closest approach corresponding to the correlation length $\xi$.
	\begin{widetext}
	\begin{center}
		\begin{table}
			\begin{tabular}{|c|c|c|c|c|c|c|c|c|c|c|c|c|}
				\hline
				&$E=140.6$ MeV	& $E=145.02$ MeV &$E=146.10$ MeV &$E=148.10$ MeV & $E=150.62$ MeV& $E=151.86$ MeV \\
				\hline
				$D_0$ (fm) &	14.8&  14.39& 14.24 &14.05  &13.81  &13.70  \\
				\hline
				$\sigma^{exp}_{1n} (\sigma^{th}_{1n})$ (mb/sr) &1.24 (1.10)	& 2.13 (2.01) & 2.32 (2.29)& 3.00 (2.96)& 3.50  (3.75)& 5.03 (4.51)\\
				\hline
				$\sigma^{exp}_{2n}$ ($\sigma^{th}_{2n}$) (mb/sr) &0.07	(0.05)& 0.23 (0.19)& 0.31 (0.26)& 0.5 (0.44)& 1.00 (0.87)& 1.83 (1.22)\\
				\hline
				\hline
				& $E=154.26$ MeV & $E=158.63$ MeV& $E=162.11$ MeV& $E=164.4$ MeV& $E=164.8$ MeV & $E=167.95$ MeV \\
				\hline
				$D_0$ (fm) &13.49  &13.12  &12.83  &12.66  & 12.62 & 12.39 \\
				\hline
				$\sigma^{exp}_{1n} (\sigma^{th}_{1n})$ (mb/sr)  & 7.25 (6.03)& 9.70  (9.08)&7.88  (9.51)& 5.92  (4.64)& 4.83  (4.53)& $<$0.7 (0.25) \\
				\hline
				$\sigma^{exp}_{2n}$ ($\sigma^{th}_{2n}$) (mb/sr)  & 2.58 (2.35)& 6.80 (7.54)& 6.11 (8.85)& 4.08 (2.34)& 3.48 (1.68)& $<$0.25 (0.07)\\
				\hline
			\end{tabular}
			\caption{Center of mass absolute differential cross section at 140$^\circ$ \cite{Montanari:14,Montanari:16,Corradi:20} associated with the reactions (\ref{eq4}). In brackets the results of the theoretical calculations carried out as explained in the text. For the twelve bombarding energies ($E=E_{c.m.}$) also the distance of closest approach $D_0$ is indicated.}\label{tab1}
		\end{table}
	\end{center}
\end{widetext}
Making the ansatz that $\theta_{c.m.}=0^\circ (\hat k_i=\hat k_f=\hat z)$,  $m_A\approx m_B, m_b\approx m_a\gg1$, as well as that the effective formfactor is constant within the overlap region and zero outside, and substituting the distorted waves by plane waves one obtains $T\sim\frac{1}{q}\sin(q\xi)\; (q=k_f-k_i)$. For small momentum transfer $(q\to0)$, $T\sim\exp\left(-(Q-E_\gamma)^2/\Delta E^2\right)$, and the FWHM of the line shape is $\Delta E\approx\sqrt{3}(\hbar/\tau_{coll})$ ($\approx$ 2.30 MeV). It is of notice that in a Josephson junction, the fact that the  two superconductors $S_\ell$ and $S_r$ are macroscopic objects at rest, implies that the delta function in (\ref{eq156})   is replaced by $\delta(\omega-2eV/h)$ which in the nuclear case translates into $\delta(E_\gamma-Q_{2n})$.

\begin{figure}[h]
	\centerline{\includegraphics*[width=9.5cm,angle=0]{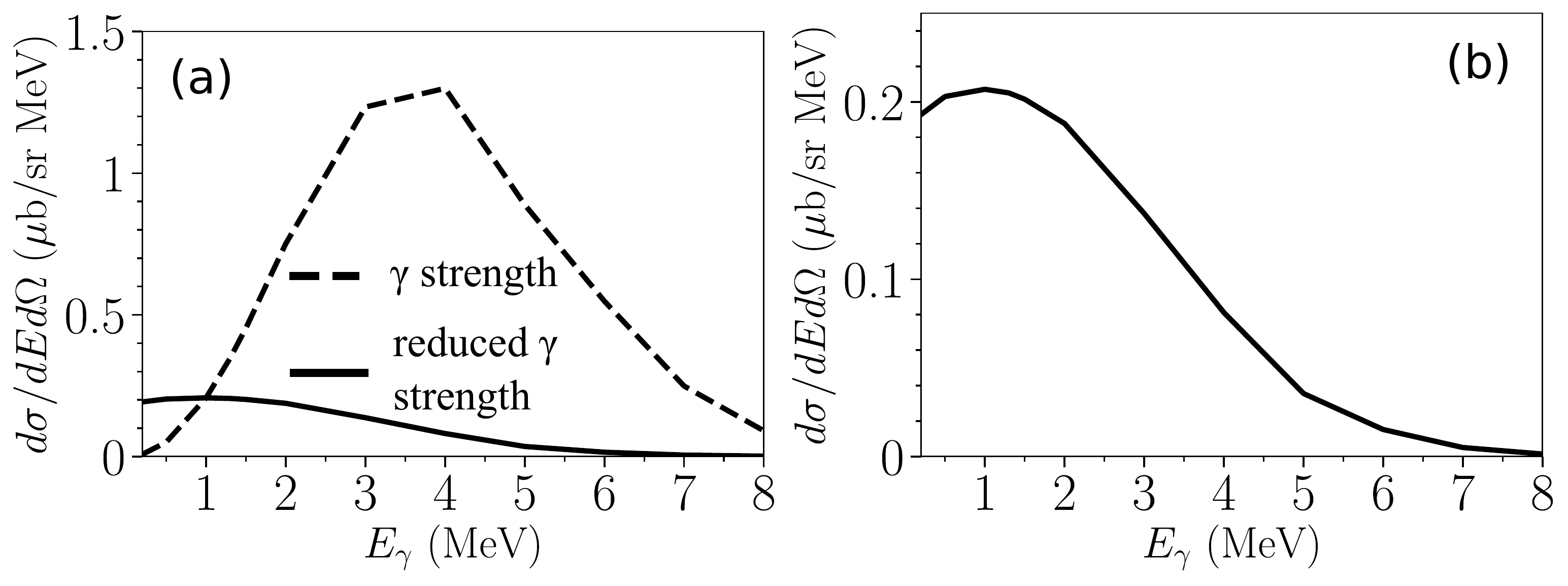}}
	\caption{(a) Double differential cross section for $\gamma$-emission at $\theta_{c.m.}=140^\circ$ as a function of the energy of the emitted $\gamma$-ray, calculated with Eqs. (\ref{eqgamma}) and (\ref{eq156}) (dashed curve). The reduced strength (continuous curve) has been obtained by dividing out from $d^2\sigma/d\Omega dE_\gamma$ the phase-space factor $\sim E_\gamma^2$, and multiplying it  by the corresponding quantity with $E_\gamma=1.307$ (MeV) (see text for details). The reduced $\gamma$-strength function is shown in (b) with a different scale, where the width and the position of the centroid are more apparent.}\label{fig4}
\end{figure}

The $\gamma$-strength function (\ref{eq156}) was worked out making use of microscopic formfactors (see Eq. (\ref{eqgamma})). They were obtained from the coherent summation of products of single-particle wavefunctions weighted by the two-nucleon spectroscopic amplitudes. These wavefunctions were calculated with the help of the mean field potentials $U^{(i)}$, potentials which also act in the transfer process, propagated from the initial to the final channel by the Green's function. The distorted waves $\chi$ were determined with the help of the microscopic optical potential of ref. \cite{Montanari:14}. Up to 150 partial waves were included in the calculation. The final results are shown in Fig. \ref{fig4} (a) in terms of a dashed line.  It describes a $\gamma$-strength function with centroid, FWHM and energy integrated area of 4 MeV, 5 MeV and 5.18 $\mu$b/sr respectively, associated with a dipole moment $\langle  d\rangle=-e\times9.36$ fm ($\langle r\rangle\approx10.52$ fm).

Multiplying these results by $\left(\frac{8\pi}{3}\frac{(1.307)^2\text{MeV}^2}{(\hbar c)^3}\right)\left(\frac{8\pi}{3}\frac{E_\gamma^2}{(\hbar c)^3}\right)^{-1}$ one obtains a Gaussian-like reduced $\gamma$-strength function (Fig. \ref{fig4} (a) and (b) continuous line). The associated centroid, FWHM and energy integrated area being: 1.1 MeV; 3.6 MeV and 0.57 $\mu$b/sr  respectively, the value of the associated dipole moment being $d=-e\times 9.36$ fm ($r=10.52$ fm). Quantities which can be compared at profit with the corresponding results of the macroscopic calculations\footnote{Making use of the value $d=-e\times9.36$ fm resulting from (\ref{eqgamma}), one can estimate, from the macroscopic prediction, a microscopic one. Namely, ($9.36/12.01)^2\times0.96\;\mu$b/sr $\approx0.58\;\mu$b/sr. A result which testifies to the validity of the separability between the $\gamma$-process (number of photons $\mathcal N$) and the two-nucleon transfer one ($\sigma_{2n}$), assumed in the macroscopic model.}.  
Summing up,  both the centroid,  width as well as the line shape of the $\gamma$-strength function are distorted as compared to the simple dipole macroscopic estimate, let alone in relation to that  observed in the radiofrequency emission from a Josephson junction (see e.g. \cite{Lindelof:81}). All this without jeopardizing the validity of the nuclear analogy.

\emph{Conclusions}---
The special effects found in superconductivity		by which a dc voltage $V$ applied across a junction between two superconductors does not determine the intensity of the supercurrent (Ohm's law) circulating through it, but the frequency  of an alternating supercurrent ($\nu_J=2e\times V/h$), finds its nuclear analogue in the electromagnetic radiation predicted to be emitted in a quasielastic heavy ion collision between two superfluid nuclei in terms of $\gamma$-rays of frequency $\nu=Q_{2n}/h$. For the particular reaction studied, and selecting the bombarding energy for which the distance of closest approach is approximately equal to the correlation length $\xi\approx13.5$ fm (largest of the measured distances of closest approach for which $\sigma_{2n}\approx \sigma_{1n}$ within a factor of two), theory predicts  for the reduced $\gamma$-strength function $d\sigma_\gamma/d\Omega|_{\theta_{c.m.}=140^\circ}\approx 0.57\, \mu$b/sr ($\nu_J\approx1.1$ MeV$/h$) corresponding to an observable $\gamma$-strength function $d\sigma_\gamma/d\Omega|_{\theta_{c.m.}=140^\circ}\approx 5.18\, \mu$b/sr, peaked at $\approx4 $ MeV. It can be concluded that a nuclear analogue to the (ac) Josephson effect has been identified. 

RAB wants to acknowledge illuminating discussions with Lorenzo Corradi and Suzana Szilner in connection with an extremely fruitful visit to the Laboratori Nazionali di Legnaro. He is also beholden to Giovanni Pollarolo concerning clarifications of the theoretical analysis, and to Chris Pethick for inspiring discussions.  This work was performed under the
auspices of the U.S. Department of Energy by Lawrence Livermore National Laboratory under Contract No. DE-AC52-07NA27344. F.B. thanks the Spanish Ministerio de Econom\'ia y Competitividad and FEDER funds under project FIS2017-88410-P.

 \end{document}